\documentclass[preprint,12pt,nopreprintline]{elsarticle}
\usepackage{amssymb}
\usepackage[para,online,flushleft]{threeparttable}
\graphicspath{ {./figures/} }

\usepackage{float}
\usepackage{verbatim}
\restylefloat{figure}
\restylefloat{table}
\usepackage{pgfplots}
\usepackage{titlesec}
\titleclass{\subsubsubsection}{straight}[\subsection]
\newcounter{subsubsubsection}[subsubsection]
\renewcommand\thesubsubsubsection{\thesubsubsection.\arabic{subsubsubsection}}
\titleformat{\subsubsubsection}{\normalfont\normalsize\itshape}{\thesubsubsubsection.\space}{0em}{}
\titlespacing*{\subsubsubsection}{0pt}{2ex plus 1ex minus .2ex}{0.75ex plus .2ex}
\makeatletter
\def\toclevel@subsubsubsection{4}
\def\l@subsubsubsection{\@dottedtocline{4}{7em}{4em}}
\makeatother

\usepackage{graphicx}
\usepackage{amsmath,amssymb,amsfonts}
\usepackage{mleftright}
\mleftright
\usepackage{algorithmic}
\usepackage{textcomp}
\usepackage{booktabs}
\usepackage{array}
\usepackage{multirow}
\usepackage{url}
\usepackage{float}
\usepackage{pifont}

\newtheorem{Proposition}{Proposition}

\newtheorem{lemma}{Lemma}
\newproof{proof}{Proof}
\floatstyle{plaintop}
\restylefloat{table}
\usepackage{caption}
\usepackage{subcaption}

\usepackage{booktabs}% http://ctan.org/pkg/booktabs

\newif\ifblackandwhite
\blackandwhitetrue
\usepackage{pdflscape}
\usepackage{colortbl}

\usepackage{booktabs}

\usepackage{adjustbox}
\usepackage{rotating}

\def\BibTeX{{\rm B\kern-.05em{\sc i\kern-.025em b}\kern-.08em
    T\kern-.1667em\lower.7ex\hbox{E}\kern-.125emX}}

\usepackage{geometry}
\begin{document}
\begin{frontmatter}

\title{ Simulation in discrete choice models evaluation: SDCM, a simulation tool for performance evaluation of DCMs}

% , Abolfazl Razi$^{b}$

\author{Amirreza Talebi$^a$}

\affiliation{organization={Department of Integrated Systems Engineering},
            addressline={The Ohio State University},
            city={Columbus},
            postcode={43210},
            state={OH},
            country={USA}}

\begin{abstract}
Discrete choice models are widely used in economics, transportation, and marketing to explain and predict how decision makers choose among a finite set of alternatives. The resulting choices depend on the socioeconomic characteristics of the decision makers, the attributes of the alternatives, and the conditions under which the decision is made. Because these factors interact in complex ways, and because real experiments raise confidentiality concerns, take time, and are costly, controlled experimentation with real respondents is often impractical. Simulation offers an alternative in which the data-generating process, and hence the true parameters, are known. This paper connects discrete choice modeling, experimental design, and simulation. We review the three areas in a common notation and state the results on which the simulation approach relies. We then formalize a simulation tool that combines an experimental design with a behavioral model of artificial agents to generate synthetic stated choice data, and we prove that the generated choices follow a panel mixed multinomial logit model with known parameters. On this basis we define measures of estimator performance and of the discrepancy between simulated and real data, illustrate the tool with a stated choice experiment on the willingness to pay for environmental attributes of roses, and discuss the behavioral factors that separate simulated choices from real ones.
\end{abstract}


\end{frontmatter}

\allowdisplaybreaks

\section{Introduction}\label{intro}

Discrete choice models (DCMs) describe how a decision maker selects one alternative from a finite set of mutually exclusive alternatives, and they do so on the basis of an explicit behavioral theory. The dominant theory is random utility maximization. Each alternative yields a utility that the analyst observes only in part, and the decision maker chooses the alternative with the highest utility. Models built on this premise are called random utility models (RUMs), and their theoretical foundations are well documented \cite{ben1999discrete, mcfadden1974utd, mcfadden2001e, train2009dc}. More recently, random regret minimization (RRM) models have offered a behavioral alternative in which decision makers try to avoid situations where a foregone alternative outperforms the chosen one on some attributes \cite{chorus2008random, chorus2010new}. Both families are now applied in transportation, marketing, environmental economics, and many other fields.

The multinomial logit (MNL) model is the workhorse among RUMs because its choice probabilities have a closed form. This tractability comes at a price. The MNL model implies the independence of irrelevant alternatives (IIA) property, and it cannot represent tastes that vary randomly across decision makers. Models that relax these restrictions, such as the mixed multinomial logit (MMNL) and the multinomial probit models, have choice probabilities that are multidimensional integrals without a closed form. Simulation-based estimation has made such models usable in practice \cite{mcfadden2001e, trainmcfadden, train2009dc}, and they are now routinely used to measure consumer preferences and the willingness to pay (WTP) for product and policy attributes.

Estimating these models requires choice data. Revealed preference data record actual market choices, but they often contain little variation in the attributes of interest and cannot describe products or policies that do not yet exist. Stated choice (SC) experiments fill this gap. Respondents face a sequence of hypothetical choice situations, each containing a small set of alternatives described by attributes whose levels are fixed by the analyst through an experimental design, typically a full or fractional factorial design or an efficient design \cite{bunch1996comparison, hess2014handbook, kirk2012experimental, kuhfeld2003marketing, rose2008designing}. Because the design determines how much information each choice carries about the parameters, the literature has moved from orthogonal designs toward efficient designs that minimize the asymptotic variance of the estimates of a non-linear choice model.

Real experiments nevertheless have two limitations. The first is practical. They are costly and slow, and confidentiality or ethical constraints sometimes make them impossible. The second is more fundamental. The parameters that generate real choices are never observed, and real choices are shaped by the gap between revealed and normative preferences, which reflects emotions, fairness, reciprocity, social norms, and bounded rationality \cite{carlsson2010design, kahneman2003maps}. A model estimated on real data therefore cannot be judged against a known truth, and a poor fit cannot be attributed with confidence to the model, the design, the sample size, or the behavior of the respondents. Simulated data remove this ambiguity. When choices are generated by a specified behavioral model with known parameters, the bias, precision, and predictive accuracy of an estimator can be measured directly, and the effect of each factor, whether internal to the decision maker or imposed by the environment, can be isolated by varying it while the others are held fixed.

This paper develops this idea and makes four contributions. First, it reviews RUMs and RRM models, the design of SC experiments, and the simulation of SC data in a common notation, and it states the formal results on which the simulation approach relies, including the logit form of the choice probabilities, the information matrix and the D-error of a design, and the invariance of design rankings by D-error to the units of the attributes. Second, it formalizes an agent-based simulation tool, implemented as an R package, as a data-generating process with explicit steps for the socioeconomic characteristics of the agents, their taste parameters, the unobserved part of utility, and the choices, and it proves that the generated data follow a panel MMNL model with known parameters. Third, it defines measures of estimator performance, namely the bias, the root mean squared error (RMSE), the coverage of confidence intervals, and the recovery of the WTP, together with a measure of the discrepancy between simulated and real choice data. Fourth, it illustrates the tool with the SC experiment of \cite{michaud2013willingness} on the WTP for environmental attributes of roses and discusses, in the same formal framework, the behavioral factors that separate simulated choices from real ones.

The rest of the paper is organized as follows. Section \ref{background} reviews RUMs and RRM models (Section \ref{rum}), experimental designs (Section \ref{ed}), and the simulation of SC data (Section \ref{sofscda}). Section \ref{methodanddiscuss} presents the simulator, its case study, the performance measures, and the sources of discrepancy between simulated and real data. Section \ref{conclusion} concludes and outlines directions for future research.

\section{Background and Related Literature}\label{background}

\subsection{Random Utility and Random Regret Models}\label{rum}

Consider a decision maker $n \in \{1,\dots,N\}$ who faces a choice set $\mathcal{C} = \{1,\dots,J\}$ of mutually exclusive alternatives. Under random utility maximization, alternative $j$ provides the utility
\begin{equation}\label{eq:utility}
U_{nj} = V_{nj} + \varepsilon_{nj},V_{nj} = \beta^{\top} x_{nj},
\end{equation}
where $x_{nj} \in \mathbb{R}^K$ collects the attributes of alternative $j$ and possibly their interactions with characteristics of $n$, $\beta \in \mathbb{R}^K$ is a vector of taste parameters, $V_{nj}$ is the systematic part of utility observed by the analyst, and $\varepsilon_{nj}$ is the part the analyst does not observe \cite{train2009dc}. The decision maker chooses alternative $i$ if and only if $U_{ni} > U_{nj}$ for all $j \neq i$. If $\varepsilon_n = (\varepsilon_{n1},\dots,\varepsilon_{nJ})^{\top}$ has density $f$, the probability that $n$ chooses $i$ is
\begin{align}\label{eq:rumprob}
P_{ni} = \Pr\left(\varepsilon_{nj} - \varepsilon_{ni} < V_{ni} - V_{nj}, \forall j \neq i\right)
= \int_{\mathbb{R}^J} \mathbf{1}\left\{\varepsilon_{nj} - \varepsilon_{ni} < V_{ni} - V_{nj}, \forall j \neq i\right\} f(\varepsilon_n)\,d\varepsilon_n,
\end{align}
where $\mathbf{1}\{\cdot\}$ is the indicator function. Because $f$ is a density, ties occur with probability zero, so \eqref{eq:rumprob} defines a probability distribution over $\mathcal{C}$. Different assumptions on $f$ lead to different DCMs.

Two properties of \eqref{eq:rumprob} matter for everything that follows. First, adding the same constant to every $U_{nj}$ leaves $P_{ni}$ unchanged, so only utility differences are identified. Alternative-specific constants can therefore be estimated only relative to a reference alternative, and characteristics of the decision maker that do not vary across alternatives can enter $V_{nj}$ only through interactions with alternative-specific variables. Second, multiplying every $U_{nj}$ by a constant $\lambda > 0$ also leaves $P_{ni}$ unchanged, so the scale of utility is not identified and is normalized by fixing the variance of $\varepsilon_{nj}$.

\begin{lemma}\label{lem:logit}
If $\varepsilon_{n1},\dots,\varepsilon_{nJ}$ are independent and identically distributed (i.i.d.) standard Gumbel random variables, with cumulative distribution function $F(\varepsilon) = \exp(-e^{-\varepsilon})$, then
\begin{equation}\label{eq:mnl}
P_{ni} = \frac{e^{V_{ni}}}{\sum_{j \in \mathcal{C}} e^{V_{nj}}}, i \in \mathcal{C}.
\end{equation}
\end{lemma}

\begin{proof}
Condition on $\varepsilon_{ni} = \varepsilon$. Alternative $i$ is chosen if and only if $\varepsilon_{nj} < \varepsilon + V_{ni} - V_{nj}$ for every $j \neq i$, and by independence this event has probability $\prod_{j \neq i} F(\varepsilon + V_{ni} - V_{nj})$. Integrating against the density $F'(\varepsilon) = e^{-\varepsilon}\exp(-e^{-\varepsilon})$ of $\varepsilon_{ni}$ gives
\begin{align*}
P_{ni} = \int_{-\infty}^{\infty} \exp\Bigl(-\sum_{j \neq i} e^{-(\varepsilon + V_{ni} - V_{nj})}\Bigr) e^{-\varepsilon} \exp\left(-e^{-\varepsilon}\right) d\varepsilon 
= \int_{-\infty}^{\infty} \exp\left(-e^{-\varepsilon} S_{ni}\right) e^{-\varepsilon}\,d\varepsilon, S_{ni} = \sum_{j \in \mathcal{C}} e^{V_{nj} - V_{ni}},
\end{align*}
because the term $j = i$ of $S_{ni}$ equals one and absorbs the factor $\exp(-e^{-\varepsilon})$. The substitution $t = e^{-\varepsilon}$ yields $P_{ni} = \int_0^{\infty} e^{-t S_{ni}}\,dt = 1/S_{ni}$, which is \eqref{eq:mnl}. \qed
\end{proof}

Equation \eqref{eq:mnl} is the MNL model \cite{mcfadden1974utd}. It makes two features of RUMs explicit. The odds between any two alternatives,
\begin{equation}\label{eq:iia}
\frac{P_{ni}}{P_{nk}} = \exp\left(V_{ni} - V_{nk}\right),
\end{equation}
do not depend on the other alternatives in $\mathcal{C}$, which is the IIA property. Moreover, if $\varepsilon_{nj}$ is Gumbel with scale $1/\lambda$, so that its variance is $\pi^2/(6\lambda^2)$, the same argument gives $P_{ni} = e^{\lambda V_{ni}}/\sum_{j \in \mathcal{C}} e^{\lambda V_{nj}}$. Only the product $\lambda\beta$ is identified, and the normalization $\lambda = 1$ fixes the unit of utility. Ratios of coefficients do not depend on this normalization. In particular, the WTP for a unit increase in attribute $k$ is the marginal rate of substitution between the attribute and the price,
\begin{equation}\label{eq:wtp}
\mathrm{WTP}_k = -\frac{\partial V_{nj}/\partial x_{njk}}{\partial V_{nj}/\partial \mathrm{Price}_{nj}} = -\frac{\beta_k}{\beta_{\mathrm{Price}}},
\end{equation}
which is the price change that keeps $V_{nj}$ constant when $x_{njk}$ increases by one unit and is invariant to $\lambda$. If an attribute also enters through an interaction, its marginal utility includes the interaction coefficient times the other attribute, and the WTP then depends on the level of that attribute.

RRM models replace utility maximization by regret minimization. In the formulation of \cite{chorus2010new}, the systematic regret of alternative $i$ is
\begin{equation}\label{eq:regret}
R_{ni} = \sum_{j \neq i} \sum_{k=1}^{K} \ln\left(1 + \exp\left[\beta_k \left(x_{njk} - x_{nik}\right)\right]\right),
\end{equation}
Each term increases with $\beta_k(x_{njk} - x_{nik})$. It is close to zero when $j$ is clearly worse than $i$ on attribute $k$ and grows almost linearly when $j$ is better, so the regret of $i$ comes mainly from foregone alternatives that outperform it. The decision maker minimizes the random regret $R_{ni} + \varepsilon_{ni}$, and if the $-\varepsilon_{ni}$ are i.i.d. standard Gumbel, Lemma \ref{lem:logit} applied to $-R_{ni}$ gives
\begin{equation}\label{eq:rrm}
P_{ni} = \frac{e^{-R_{ni}}}{\sum_{j \in \mathcal{C}} e^{-R_{nj}}}.
\end{equation}
Because $R_{ni}$ depends on the attributes of every alternative in $\mathcal{C}$, the ratio $P_{ni}/P_{nk} = \exp(R_{nk} - R_{ni})$ in general changes when an alternative is added or removed. The RRM model therefore does not satisfy IIA, and it can reproduce compromise effects that the MNL model cannot.

In some contexts decision makers appear to minimize regret rather than maximize utility, which calls for DCMs based on the appropriate behavioral theory \cite{hensher2013random, gusarov2020exploration}. \cite{chorus2014random} assessed the empirical evidence on regret-based and utility-based consumer choice models and found that the regret-based models often fit and predicted better, with differences that were small but large enough to matter for managerial implications. \cite{masiero2019understanding} examined the hotel location preferences of tourists with hypothetical alternatives and found RRM-based models superior to RUM-based ones. \cite{sharma2019park} modeled park-and-ride lot choice with both decision rules and showed that RRM models predicted better and captured the trade-offs between the auto and transit networks. \cite{mao2020does} analyzed public preferences for air quality policies and found that RRM models fit better and that regret-driven respondents favored cleaner air and fewer haze days, which can guide the design of effective policies. \cite{iraganaboina2021evaluating} studied route choice in the Greater Orlando region and stressed the value of customized RRM models for understanding travel behavior and designing traffic management systems. \cite{wong2020revealed} assessed evacuation behavior during the 2017 Southern California wildfires. RRM did not clearly dominate RUM in that study, owing to limited attribute variation, but weak regret aversion and class-specific regret were observed, which suggests further study of RRM models for evacuation. The rest of this paper focuses on RUMs, and RRM returns in Section \ref{conclusion} as an extension of the simulator.

The MMNL model removes the restrictions of the MNL model by letting tastes vary across decision makers. Conditional on a taste vector $\beta$, the choice probability is the logit kernel
\begin{equation}\label{eq:kernel}
L_{ni}(\beta) = \frac{\exp(\beta^{\top} x_{ni})}{\sum_{j \in \mathcal{C}} \exp(\beta^{\top} x_{nj})},
\end{equation}
and $\beta$ is distributed in the population with a density $f(\beta \mid \theta)$, usually continuous, whose parameters $\theta$ are, for instance, a mean vector and a covariance matrix. The unconditional choice probability is
\begin{equation}\label{eq:mmnl}
P_{ni} = \int_{\mathbb{R}^K} L_{ni}(\beta) f(\beta \mid \theta)\,d\beta.
\end{equation}
The ratio $P_{ni}/P_{nk}$ now depends on all alternatives, so IIA no longer holds, and \cite{trainmcfadden} showed that, under mild regularity conditions, the choice probabilities of any RUM can be approximated arbitrarily closely by an MMNL model. This flexibility, together with its straightforward application, explains why the MMNL model is regarded as one of the most effective DCMs.

In an SC experiment each respondent answers $S$ choice situations, and tastes are usually assumed to be stable across the choices of the same person. Let $y_{nsj} = 1$ if respondent $n$ chooses alternative $j$ in situation $s$ and $y_{nsj} = 0$ otherwise. The probability of the observed sequence of choices $y_n$ is then
\begin{equation}\label{eq:panel}
P_n(\theta) = \int_{\mathbb{R}^K} \prod_{s=1}^{S} \prod_{j \in \mathcal{C}} L_{nsj}(\beta)^{y_{nsj}} f(\beta \mid \theta)\,d\beta,
\end{equation}
where $L_{nsj}$ is \eqref{eq:kernel} evaluated at the attributes $x_{nsj}$ of situation $s$. Neither \eqref{eq:mmnl} nor \eqref{eq:panel} has a closed form, so both are approximated by simulation. With $R$ independent draws $\beta^{(1)},\dots,\beta^{(R)}$ from $f(\cdot \mid \theta)$,
\begin{equation}\label{eq:sim}
\check{P}_n(\theta) = \frac{1}{R} \sum_{r=1}^{R} \prod_{s=1}^{S} \prod_{j \in \mathcal{C}} L_{nsj}\bigl(\beta^{(r)}\bigr)^{y_{nsj}},
\end{equation}
and the maximum simulated likelihood (MSL) estimator maximizes $\check{\ell}(\theta) = \sum_{n=1}^{N} \ln \check{P}_n(\theta)$. The simulator $\check{P}_n$ is unbiased for $P_n$, but the logarithm is strictly concave, so Jensen's inequality gives $\mathbb{E}[\ln \check{P}_n(\theta)] < \ln P_n(\theta)$ whenever $\check{P}_n(\theta)$ is not almost surely constant, and the simulated log-likelihood is biased downward for every finite $R$. The MSL estimator is consistent when $R \to \infty$ as $N \to \infty$, and it is asymptotically equivalent to the maximum likelihood estimator when $R$ grows faster than $\sqrt{N}$ \cite{train2009dc}. Estimates obtained by simulation are therefore approximations whose accuracy is governed by $R$.

DCMs are also closely related to machine learning classifiers. When the systematic utility is $V_{nj} = \beta_j^{\top} w_n$, with alternative-specific coefficients $\beta_j$ (one of which is normalized to zero) and a vector $w_n$ of characteristics of the decision maker, \eqref{eq:mnl} is exactly the softmax, or multinomial logistic, regression of machine learning, and maximizing the log-likelihood $\sum_{n} \sum_{j} y_{nj} \ln P_{nj}$ is equivalent to minimizing the cross-entropy loss. For this reason MNL and MMNL models built on RUM theory are routinely compared with machine learning methods for forecasting choices \cite{gusarov2020exploration, hillel2021systematic, soleymani_forecasting_2024, talebi2024integrating}.

\subsection{Experimental Design}\label{ed}

An experimental design specifies the independent, dependent, and control variables of an experiment together with its randomization and statistical procedures \cite{kirk2012experimental}. In an SC experiment the design is a matrix $X$ that stacks the attribute vectors $x_{sj} \in \mathbb{R}^K$ of the alternatives $j \in \mathcal{C}$ in the choice situations $s \in \{1,\dots,S\}$. If the alternatives are described by $A$ attributes and attribute $a$ takes $\ell_a$ levels, the full factorial design contains all $\prod_{a=1}^{A} \ell_a$ profiles, and a fractional factorial design uses a subset of them. Designing an SC experiment thus amounts to filling the design matrix with attribute levels, that is, to choosing which profiles are combined into each choice situation.

Choice situations have traditionally been built from orthogonal designs \cite{rose2008stated}, in which the centered attribute columns are mutually uncorrelated and every level of an attribute appears equally often. Orthogonal designs are optimal for linear models and allow all main effects, and in a full factorial also all interactions, to be estimated independently, which explains their long use \cite{kuhfeld2003marketing}. Choice models are non-linear, however, and for them orthogonality is generally not efficient \cite{hess2014handbook}. The reason is visible in the Fisher information matrix of the MNL model. For a single respondent who answers all $S$ situations,
\begin{align}
I_1(\beta \mid X) = \sum_{s=1}^{S} \sum_{j \in \mathcal{C}} P_{sj}(\beta) \left(x_{sj} - \bar{x}_s(\beta)\right)\left(x_{sj} - \bar{x}_s(\beta)\right)^{\top},
\bar{x}_s(\beta) = \sum_{j \in \mathcal{C}} P_{sj}(\beta) x_{sj},  \label{eq:info}
\end{align}
where $P_{sj}(\beta)$ is \eqref{eq:mnl} evaluated at the attributes $x_{sj}$. This expression follows by differentiating the log-likelihood $\sum_{s} \sum_{j} y_{sj} \ln P_{sj}(\beta)$ twice, since its Hessian equals $-I_1(\beta \mid X)$ and does not depend on the choices. The asymptotic variance-covariance (AVC) matrix of the maximum likelihood estimator is $\Omega_1(X,\beta) = I_1(\beta \mid X)^{-1}$ for one respondent and $\Omega_1(X,\beta)/N$ for $N$ respondents who face the same design. Three consequences follow from \eqref{eq:info}.

First, the information depends on the unknown $\beta$ through $P_{sj}(\beta)$, so prior values of the parameters are needed to evaluate a design. Second, at $\beta = 0$ all alternatives are equally likely, $P_{sj} = 1/J$, and
\begin{equation}\label{eq:info0}
I_1(0 \mid X) = \frac{1}{J} \sum_{s=1}^{S} \sum_{j \in \mathcal{C}} \left(x_{sj} - \bar{x}_s\right)\left(x_{sj} - \bar{x}_s\right)^{\top}, \bar{x}_s = \frac{1}{J}\sum_{j \in \mathcal{C}} x_{sj},
\end{equation}
which does not depend on $\beta$ and has the form of the moment matrix of a linear model in the attributes centered within each choice situation. This is why orthogonal and level-balanced designs perform well when the parameters are close to zero \cite{street2007construction}, and why they lose efficiency when $\beta \neq 0$ makes the weights $P_{sj}(\beta)$ unequal. Third, the contribution of situation $s$ to \eqref{eq:info} is the covariance matrix of the attributes under the choice probabilities $P_{s1},\dots,P_{sJ}$. If one alternative dominates, so that $P_{si} \to 1$, then $\bar{x}_s \to x_{si}$ and this contribution vanishes. Choice situations with dominant alternatives carry almost no information, which is why efficient designs, which avoid them, produce lower standard errors than orthogonal designs \cite{bliemer2011experimental}.

Efficient designs minimize a scalar function of the AVC matrix. Researchers who focus on the standard errors of individual estimates minimize the A-error, which averages the diagonal elements of $\Omega_1$, while the most common criterion is the D-error \cite{rose2009constructing},
\begin{equation}\label{eq:derror}
\text{A-error}(X,\beta) = \frac{\operatorname{tr} \Omega_1(X,\beta)}{K}, \text{D-error}(X,\beta) = \left[\det \Omega_1(X,\beta)\right]^{1/K}.
\end{equation}
The D-error is the $K$-th root of the generalized variance of the estimates, so a lower D-error indicates a more efficient design and more precise parameter estimates. Its popularity rests partly on the following invariance property, which the A-error lacks.

\begin{Proposition}\label{prop:invariance}
Let $D = \operatorname{diag}(d_1,\dots,d_K)$ with $d_k \neq 0$, and suppose every attribute is expressed in new units, $\tilde{x}_{sj} = D x_{sj}$, with the corresponding parameters $\tilde{\beta} = D^{-1}\beta$. Then the choice probabilities are unchanged and, for every design $X$,
\begin{equation*}
\text{D-error}(\tilde{X},\tilde{\beta}) = |\det D|^{-2/K} \text{D-error}(X,\beta).
\end{equation*}
Hence the ranking of designs by D-error does not depend on the units of the attributes, whereas the ranking by A-error can.
\end{Proposition}

\begin{proof}
Since $\tilde{\beta}^{\top}\tilde{x}_{sj} = \beta^{\top} D^{-1} D x_{sj} = \beta^{\top} x_{sj}$, all utilities and choice probabilities are unchanged. The weighted means in \eqref{eq:info} become $D\bar{x}_s(\beta)$, so $I_1(\tilde{\beta} \mid \tilde{X}) = D I_1(\beta \mid X) D$ and $\Omega_1(\tilde{X},\tilde{\beta}) = D^{-1} \Omega_1(X,\beta) D^{-1}$. Therefore $\det \Omega_1(\tilde{X},\tilde{\beta}) = \det \Omega_1(X,\beta)/(\det D)^2$, and taking the power $1/K$ proves the identity. The factor $|\det D|^{-2/K}$ is common to all designs, so the ranking is preserved. For the A-error, $\operatorname{tr} \Omega_1(\tilde{X},\tilde{\beta}) = \sum_{k} [\Omega_1(X,\beta)]_{kk}/d_k^2$ weights the variances by the units. For instance, two designs with $\Omega_1 = \operatorname{diag}(1,3)$ and $\Omega_1 = \operatorname{diag}(3,2)$ have traces 4 and 5, but after rescaling with $D = \operatorname{diag}(1,1/2)$ their traces become 13 and 11, so their order is reversed. \qed
\end{proof}

Rather than relying on fixed prior values, Bayesian efficient designs describe the prior knowledge by a density $\pi(\beta)$ and minimize the expected D-error
\begin{equation}\label{eq:dberror}
\text{D}_b\text{-error}(X) = \int_{\mathbb{R}^K} \left[\det \Omega_1(X,\beta)\right]^{1/K} \pi(\beta)\,d\beta,
\end{equation}
which in practice is approximated by averaging over draws from $\pi$. This makes the design robust to misspecified priors \cite{bliemer2011experimental, rose2009constructing, rose2008stated}. Even minimal prior information, such as the signs of the parameters, can improve efficiency, but accurate priors remain important, and \cite{bliemer2009efficient} showed that misspecified priors can produce a less efficient design than zero priors.

The information matrix also links the design to the sample size. With $N$ respondents, the asymptotic $t$-ratio of $\hat{\beta}_k$ is $|\beta_k|/\sqrt{[\Omega_1(X,\beta)]_{kk}/N}$, and it is at least a critical value $t_{\alpha}$ if and only if
\begin{equation}\label{eq:samplesize}
N \geq N_k^{\ast} = \frac{t_{\alpha}^2 [\Omega_1(X,\beta)]_{kk}}{\beta_k^2}.
\end{equation}
A design therefore needs at least $\max_k N_k^{\ast}$ respondents for all parameters to be statistically significant, and this theoretical minimum sample size can be computed for each parameter before data are collected \cite{rose2009constructing}. Since $N_k^{\ast}$ is proportional to $[\Omega_1]_{kk}$, a design that halves a variance also halves the required sample, so design efficiency can matter as much as the number of respondents.

When constructing SC designs, labeling, parameter types, attribute levels, and interaction terms must also be considered. Preference heterogeneity and interaction effects can substantially change design efficiency and must be built into the design \cite{bliemer2010construction}, and the range of continuous attributes and the selection of choice sets affect both the efficiency and the practical relevance of the experiment \cite{hess2014handbook}.

Efficient designs are found by search. The modified Fedorov algorithm \cite{rose2008designing, street2005quick, voytekhovsky2001fedorov} starts from a design $X^{(0)}$ with a nonsingular information matrix, whose rows are drawn from a candidate set $\mathcal{X}$, typically the full factorial. At iteration $t$ it evaluates every design obtained from $X^{(t)}$ by exchanging one row with one candidate in $\mathcal{X}$, and it moves to the best of them if that design lowers the criterion, for instance \eqref{eq:derror} or \eqref{eq:dberror}. Because the criterion decreases strictly at every move and only finitely many designs of a given size can be formed from a finite $\mathcal{X}$, the algorithm stops after finitely many iterations at a design that no single exchange can improve. Such a design is locally but not necessarily globally optimal, so the search is usually restarted from several initial designs. These methods illustrate the continuing refinement of experimental design in SC research.

\subsection{Simulation of SC Data}\label{sofscda}

DCMs are used extensively in fields such as transportation and marketing to explain and predict decisions among alternatives. The rationale is that a model estimated from SC data can predict the choices of agents, and the data needed for estimation can come from revealed preferences, surveys, or simulation. As discussed in Section \ref{intro}, collecting revealed preferences or running surveys is not always feasible or desirable because of time, cost, ethical, and confidentiality issues. Moreover, researchers often want to study one factor of the decision process in isolation, which a real controlled experiment can almost never achieve. Simulation can play a central role here, and since the advent of computer-aided simulation a large body of work has applied it.

Agent-based simulation, which models consumers' purchase behavior as the outcome of individual decisions, has gained particular popularity. Agent-based modeling defines agents as autonomous decision makers, each of which evaluates its situation and acts according to a predefined set of rules \cite{bonabeau2002agent, gilbert2019agent}. In the discrete choice setting, agent $n$ is described by a state $(z_n, \beta_n)$, where $z_n$ contains its socioeconomic characteristics and $\beta_n$ its tastes, and its rule in choice situation $s$ is
\begin{equation}\label{eq:abm}
a_{ns} = \arg\max_{j \in \mathcal{C}} \left( V_{nsj} + \gamma m_{-n,sj} + \varepsilon_{nsj} \right),
\end{equation}
where $m_{-n,sj}$ is the share of the other agents, or of the peers of $n$, who choose $j$, and $\gamma \geq 0$ measures social influence. With $\gamma = 0$ the agents decide independently and \eqref{eq:abm} reduces to random utility maximization. With $\gamma > 0$ the rule represents conformity or herd behavior. When the errors are Gumbel and agents respond to the expected shares $m_{sj}$ of the whole population, these shares must satisfy the fixed-point condition
\begin{equation}\label{eq:fixedpoint}
m_{sj} = \frac{1}{N}\sum_{n=1}^{N} \frac{\exp\left(V_{nsj} + \gamma m_{sj}\right)}{\sum_{k \in \mathcal{C}} \exp\left(V_{nsk} + \gamma m_{sk}\right)}, j \in \mathcal{C}.
\end{equation}
The right side maps the probability simplex continuously into itself, so a solution exists by Brouwer's fixed-point theorem, but it need not be unique. If all $V_{nsj}$ are equal, $m_{sj} = 1/J$ always solves \eqref{eq:fixedpoint}, and the linearization of the right side at this point has slope $\gamma/J$ on the simplex, so a second, asymmetric equilibrium exists whenever $\gamma > J$. \cite{brock2003mcsi} give general conditions under which such multiple self-consistent equilibria arise.

\cite{zhang2019agent} investigated the factors that influence consumer purchase behavior with an agent-based simulation built around a utility function. Taking quality, price, and promotion as the main drivers of choice, they implemented the simulation in NetLogo and analyzed the effect of each factor. \cite{zhang2007agent} drew on psychology, marketing, sociology, and engineering, the main fields that shape the study of consumer behavior, and developed an agent-based model whose motivation function combines psychological personality traits with interactions in a competitive market, which allowed them to reproduce the decoy effect. Generating heterogeneous artificial consumers within a simulated market made it possible to handle the dynamics and complications observed in real settings.

\cite{lockshin2006using} studied consumers' wine choices and used simulations based on a discrete choice experiment to examine how purchase rates change with brand, region, and awards. Respondents chose among proposed alternatives, and their choices were converted into utilities with an MNL model. \cite{munizaga2005testing} tested mixed logit and probit models in a controlled setting with synthetic data obtained by simulating the utility-maximizing choices of hypothetical individuals. They examined how the number of simulation draws affects the recovery of a correlated error structure, and they studied the use of Halton sequences in model calibration. Halton sequences matter because the accuracy of \eqref{eq:sim} depends on how evenly the draws cover the domain of integration. With independent pseudo-random draws, the standard deviation of the simulation error in \eqref{eq:sim} is of order $R^{-1/2}$, whereas a Halton sequence in $d$ dimensions has a star discrepancy of order $(\ln R)^d/R$, so by the Koksma-Hlawka inequality the integration error for integrands of bounded variation decreases almost at rate $R^{-1}$. In MSL estimation, Halton draws therefore reach a given accuracy with far fewer draws than pseudo-random sequences \cite{train2009dc}.

Simulated data also have drawbacks. Simulation provides approximations of estimates rather than exact estimates, as the downward bias of the simulated log-likelihood in Section \ref{rum} shows, and any quantity averaged over $M$ independent replications of a simulation carries a Monte Carlo standard error of order $M^{-1/2}$. A second issue concerns human nature. Simulated agents are usually assumed to be fully rational, but the complexity of human psychology often leads to irrational choices in reality, and individual decisions can be influenced by herd behavior, which corresponds to $\gamma > 0$ in \eqref{eq:abm}. To understand these sources of bias, one has to study the factors that influence decision making. \cite{wendel2020designing} identified limited time, attention, willpower, and experience, the interplay of the conscious and unconscious mind, and heuristics as major influences on decisions. Measuring these latent factors is very difficult in real experiments and even harder in simulation.

\section{Methodology and Discussion}\label{methodanddiscuss}

\subsection{The Simulator}\label{simulator}

This section details the R package SDCM (simulation of discrete choice models) developed by \cite{Amirreza2019}. R is a widely used, open-source, and platform-independent language with a large collection of packages contributed by researchers and programmers. To the best of our knowledge, no other R package is designed specifically for the simulation of SC data for DCMs. The package, which is still under development, generates simulated data for evaluating the performance of DCMs. Its open-source nature allows researchers to adapt and extend it to new scenarios, so that it can grow into a more versatile and sophisticated tool for generating controlled data. This is particularly valuable when real experiments are costly or impractical. With the tool, researchers can run preliminary studies before an actual experiment and plan more precisely the number of respondents, the choice of DCM, and other aspects of the study. The tool can also be used to compare the performance of DCMs with that of artificial neural networks and other classifiers.

The simulator generates data for $N$ artificial respondents who face the $S$ choice situations of a design $X$, with $J$ alternatives per situation, possibly including an opt-out alternative. Its inputs are the design, the marginal distributions $F_1,\dots,F_Q$ of $Q$ socioeconomic characteristics, the mean vector $\mu \in \mathbb{R}^K$ and the covariance matrix $\Sigma$ of the parameters, and the distribution of the unobserved part of utility. The data are generated in four steps.
\begin{enumerate}
\item \label{step:socio} The socioeconomic characteristics $z_n = (z_{n1},\dots,z_{nQ})^{\top}$ of each respondent are drawn independently across respondents and characteristics, with $z_{nq} \sim F_q$.
\item \label{step:taste} The parameters of each respondent are drawn once and kept for all $S$ situations,
\begin{equation}\label{eq:beta}
\beta_n = \mu + L\xi_n, \xi_n \sim \mathcal{N}(0, I_K), L L^{\top} = \Sigma,
\end{equation}
where $L$ is the lower-triangular Cholesky factor of $\Sigma$. Since $\beta_n$ is an affine function of a Gaussian vector, $\beta_n \sim \mathcal{N}(\mu, L\mathbb{E}[\xi_n \xi_n^{\top}]L^{\top}) = \mathcal{N}(\mu, \Sigma)$. Parameters that are fixed in the population correspond to zero rows and columns of $\Sigma$, and $L$ is then the Cholesky factor of the block of random parameters, padded with zeros.
\item \label{step:utility} The utility of alternative $j$ in situation $s$ is $U_{nsj} = V_{nsj} + \varepsilon_{nsj}$ with $V_{nsj} = \beta_n^{\top} x_{nsj}$, where $x_{nsj}$ contains the design attributes of $(s,j)$, their interactions, and the interactions of $z_n$ with alternative-specific dummies. The errors $\varepsilon_{nsj}$ are i.i.d. standard Gumbel. They can be generated as $\varepsilon_{nsj} = -\ln(-\ln u_{nsj})$ with $u_{nsj}$ i.i.d. uniform on $(0,1)$, because $F^{-1}(u) = -\ln(-\ln u)$ for $F(\varepsilon) = \exp(-e^{-\varepsilon})$.
\item \label{step:choice} Each respondent chooses the alternative with the highest utility,
\begin{equation}\label{eq:choice}
y_{nsj} = \mathbf{1}\Bigl\{U_{nsj} = \max_{k \in \mathcal{C}} U_{nsk}\Bigr\}.
\end{equation}
\end{enumerate}
The output is the data set $\mathcal{D}_N = \{(X, z_n, y_n)\}_{n=1}^{N}$, where $y_n$ collects the indicators $y_{nsj}$, and repeating the four steps with independent random numbers produces independent replications.

\begin{Proposition}\label{prop:dgp}
Let the data be generated by Steps \ref{step:socio} to \ref{step:choice}. Then the choices are independent across respondents and, if $\Sigma$ is positive definite, the choices of respondent $n$ have, conditional on the design and on $z_n$, the probability
\begin{equation}\label{eq:dgp}
\Pr\left(y_n \mid X, z_n\right) = \int_{\mathbb{R}^K} \prod_{s=1}^{S} \prod_{j \in \mathcal{C}} L_{nsj}(\beta)^{y_{nsj}} \phi_K(\beta \mid \mu, \Sigma)\,d\beta,
\end{equation}
where $L_{nsj}(\beta) = \exp(\beta^{\top} x_{nsj})/\sum_{k \in \mathcal{C}} \exp(\beta^{\top} x_{nsk})$ and $\phi_K(\cdot \mid \mu,\Sigma)$ is the $K$-variate normal density. The simulated data thus follow the panel MMNL model \eqref{eq:panel} with $f(\beta \mid \theta) = \phi_K(\beta \mid \mu, \Sigma)$ and true parameters $\theta^0 = (\mu, \Sigma)$. If some parameters are fixed, so that $\Sigma$ is singular, the same holds with the integral taken over the random parameters only, and if $\Sigma = 0$ the choices are independent across situations with the MNL probabilities $L_{nsj}(\mu)$.
\end{Proposition}

\begin{proof}
Fix $n$ and condition on $\beta_n = \beta$. The error vectors $(\varepsilon_{ns1},\dots,\varepsilon_{nsJ})$ are independent across $s$ and independent of $\beta_n$ and $z_n$, so the choices in different situations are conditionally independent, and by Lemma \ref{lem:logit} the choice in situation $s$ is $j$ with probability $L_{nsj}(\beta)$. Hence $\Pr(y_n \mid X, z_n, \beta_n = \beta) = \prod_{s} \prod_{j} L_{nsj}(\beta)^{y_{nsj}}$. By Step \ref{step:taste}, $\beta_n \sim \mathcal{N}(\mu,\Sigma)$ independently of $z_n$ and of the errors, and integrating over this distribution gives \eqref{eq:dgp}. Independence across respondents holds because $z_n$, $\xi_n$, and the errors are drawn independently for each $n$. With fixed parameters, $\beta_n$ has a normal distribution on the coordinates of the random parameters only, and the same argument applies. If $\Sigma = 0$, then $\beta_n = \mu$ almost surely and \eqref{eq:dgp} reduces to $\prod_{s} \prod_{j} L_{nsj}(\mu)^{y_{nsj}}$. \qed
\end{proof}

Proposition \ref{prop:dgp} is what makes the simulator a tool for performance evaluation. Because the generating model and $\theta^0$ are known, any estimator applied to $\mathcal{D}_N$ can be compared with the truth. If the estimated model is the panel MMNL with normal mixing, it is correctly specified, and under standard regularity conditions the MSL estimator satisfies $\hat{\theta} \to \theta^0$ in probability and $\sqrt{N}(\hat{\theta} - \theta^0) \to \mathcal{N}(0, I(\theta^0)^{-1})$ in distribution as $N \to \infty$ and $R/\sqrt{N} \to \infty$ \cite{train2009dc}, where $I(\theta^0)$ is the information matrix of one respondent. A systematic gap between $\hat{\theta}$ and $\theta^0$ in a simulation study can then be traced to the finite sample, the number of draws, or the design. If the estimated model is misspecified relative to the generating one, for instance an MNL model fitted to data generated with $\Sigma \neq 0$, the maximum likelihood estimator converges to the pseudo-true value that minimizes the Kullback-Leibler divergence of the estimated model from the generating one \cite{white1982maximum}, and the simulator shows how far this value lies from $\mu$. Other DCMs are obtained by changing a single step. Normally distributed errors in Step \ref{step:utility} produce a multinomial probit model, and replacing $V_{nsj}$ by the negative regret $-R_{nsj}$ of \eqref{eq:regret} produces an RRM model.

\subsection{Case Study}\label{casestudy}

The tool is illustrated with the real choice experiment of \cite{michaud2013willingness}, which examined consumers' WTP and the price premium for two environmental attributes of roses. Each choice situation contains two unlabeled alternatives, Rose A and Rose B, and an opt-out alternative, so that $\mathcal{C} = \{A, B, 0\}$ and $J = 3$. The roses are described by two binary attributes, $\mathrm{Label}_{sj} \in \{0,1\}$ and $\mathrm{Carbon}_{sj} \in \{0,1\}$, and by a price. The binary attributes yield $2 \times 2 = 4$ profiles and $\binom{4}{2} = 6$ distinct pairs of profiles. Each pair is presented twice, once in each order, which gives $S = 12$ choice situations, or 24 rose profiles, per respondent, and each profile appears equally often in the first and the second position. Prices are drawn independently and uniformly from the seven levels $\{1.5, 2, 2.5, 3, 3.5, 4, 4.5\}$ and assigned to the profiles. The simulator rebuilds this design from the full factorial of the two binary attributes. It also accepts any other design, provided that the order of the prior parameters matches the order of the design columns, and orthogonal, full or fractional factorial, and D-efficient designs, including designs with high-order interactions, can be constructed with existing R packages \cite{gromping2018r, gronmping2014r, traetsgenerating}.

The systematic utilities are
\begin{align}\label{eq:casevu}
V_{nsj} &= \mathrm{BUY}_j \left( \alpha_n + \beta_{\mathrm{Sex}}\mathrm{Sex}_n + \beta_{\mathrm{Age}}\mathrm{Age}_n + \beta_{\mathrm{Inc}}\mathrm{Income}_n + \beta_{\mathrm{Hab}}\mathrm{Habit}_n \right) + \beta_{\mathrm{Price}}\mathrm{Price}_{sj} \nonumber\\
&\quad + \beta_{n,\mathrm{L}}\mathrm{Label}_{sj} + \beta_{n,\mathrm{C}}\mathrm{Carbon}_{sj} + \beta_{n,\mathrm{LC}}\mathrm{Label}_{sj}\mathrm{Carbon}_{sj}, j \in \mathcal{C},
\end{align}
where $\mathrm{BUY}_j = 1$ for $j \in \{A, B\}$ and $\mathrm{BUY}_0 = 0$, $\mathrm{Habit}_n$ indicates organic purchase habits, and all attributes of the opt-out alternative are zero, so that $V_{ns0} = 0$. The alternative-specific constant $\alpha_n$ is the coefficient of $\mathrm{BUY}_j$ and measures the utility of buying a rose relative to buying nothing. There is no brand distinction between Rose A and Rose B, so choices between them depend only on the attributes, and the two roses share one constant, which captures the decision to buy. The characteristics in $z_n$ do not vary across alternatives, so, as shown in Section \ref{rum}, they enter only through their interaction with $\mathrm{BUY}_j$. Following the case study, $\alpha_n$, $\beta_{n,\mathrm{L}}$, $\beta_{n,\mathrm{C}}$, and $\beta_{n,\mathrm{LC}}$ are random and drawn as in \eqref{eq:beta}, while $\beta_{\mathrm{Price}}$ and the coefficients $\beta_{\mathrm{Sex}}$, $\beta_{\mathrm{Age}}$, $\beta_{\mathrm{Inc}}$, and $\beta_{\mathrm{Hab}}$ of the characteristics are fixed, so that $K = 9$ and four of the nine parameters are random. All values are taken from the estimates of \cite{michaud2013willingness}.

The fixed price coefficient gives the WTP a simple distribution. Because of the interaction term, the marginal utility of the label is $\beta_{n,\mathrm{L}} + \beta_{n,\mathrm{LC}}\mathrm{Carbon}_{sj}$. By \eqref{eq:wtp}, the WTP of respondent $n$ for the label on a rose without the carbon attribute is therefore $\mathrm{WTP}_{n,\mathrm{L}} = -\beta_{n,\mathrm{L}}/\beta_{\mathrm{Price}}$, and if $\beta_{n,\mathrm{L}} \sim \mathcal{N}(\mu_{\mathrm{L}},\sigma_{\mathrm{L}}^2)$, then
\begin{equation}\label{eq:wtpdist}
\mathrm{WTP}_{n,\mathrm{L}} \sim \mathcal{N}\left( -\frac{\mu_{\mathrm{L}}}{\beta_{\mathrm{Price}}}, \frac{\sigma_{\mathrm{L}}^2}{\beta_{\mathrm{Price}}^2} \right).
\end{equation}
The premium for a rose with both attributes over a rose with neither is also normal,
\begin{equation}\label{eq:premium}
-\frac{\beta_{n,\mathrm{L}} + \beta_{n,\mathrm{C}} + \beta_{n,\mathrm{LC}}}{\beta_{\mathrm{Price}}} \sim \mathcal{N}\left( -\frac{\mu_{\mathrm{L}} + \mu_{\mathrm{C}} + \mu_{\mathrm{LC}}}{\beta_{\mathrm{Price}}}, \frac{\iota^{\top}\Sigma_{\mathrm{env}}\iota}{\beta_{\mathrm{Price}}^2} \right),
\end{equation}
where $\Sigma_{\mathrm{env}}$ is the covariance matrix of $(\beta_{n,\mathrm{L}}, \beta_{n,\mathrm{C}}, \beta_{n,\mathrm{LC}})$ and $\iota = (1,1,1)^{\top}$. If the price coefficient were normally distributed as well, the WTP would be a ratio of normal variables whose denominator has a positive density at zero, and such a ratio generally has no finite mean. Fixing $\beta_{\mathrm{Price}}$ keeps the moments of the WTP well defined.

\cite{michaud2013willingness} also considered four socioeconomic characteristics, namely sex, age, income, and organic purchase habits. No information on their correlations was available, so they are generated independently, as in Step \ref{step:socio}. The binary characteristics are drawn by inverse transform sampling. In general, a characteristic with distribution function $F_q$ is generated as
\begin{equation}\label{eq:inverse}
z_{nq} = F_q^{-1}(u_{nq}), F_q^{-1}(u) = \inf\{ z : F_q(z) \geq u \}, u_{nq} \sim \mathcal{U}(0,1),
\end{equation}
and $\Pr(z_{nq} \leq z) = \Pr(u_{nq} \leq F_q(z)) = F_q(z)$, so $z_{nq}$ has the target distribution whether it is discrete or continuous. For sex, with a share of 0.49 of women in the case study, \eqref{eq:inverse} becomes
\begin{equation}\label{eq:sex}
\mathrm{Sex}_n = \mathbf{1}\{u_n > 0.49\},
\end{equation}
so a respondent is female ($\mathrm{Sex}_n = 0$) if $u_n$ falls in $(0, 0.49]$, which happens with probability 0.49. This construction yields exactly a Bernoulli variable with success probability 0.51, so it generates data with the same distribution as a direct Bernoulli draw. Its advantage is generality, since \eqref{eq:inverse} applies unchanged to any discrete or continuous characteristic. The habit characteristic is generated in the same way with its own probability.

Age and income are drawn from normal distributions with the means and standard deviations of the case study, truncated to admissible ranges by rejection sampling. A candidate $w = \nu + \tau\zeta$ with $\zeta \sim \mathcal{N}(0,1)$ is accepted if $a \leq w \leq b$ and redrawn otherwise. Each candidate is accepted with probability $\kappa = \Phi((b-\nu)/\tau) - \Phi((a-\nu)/\tau)$, where $\Phi$ and $\phi$ are the standard normal distribution and density functions, so $1/\kappa$ candidates are needed on average, and the accepted values have the truncated normal density
\begin{equation}\label{eq:truncnorm}
g(w) = \frac{\phi\left((w - \nu)/\tau\right)}{\tau\kappa}, a \leq w \leq b.
\end{equation}
Age is truncated between 18 and 85 years. Income is truncated below at zero, so that $\kappa = 1 - \Phi(-\nu/\tau)$, and grouped into six classes $[c_{h-1}, c_h)$, $h = 1,\dots,6$, with $c_0 = 0$ and $c_6 = \infty$, so that class $h$ has probability $\rho_h = [\Phi((c_h - \nu)/\tau) - \Phi((c_{h-1} - \nu)/\tau)]/\kappa$.

Independence is an assumption forced by missing information rather than a property of real populations, and correlated characteristics, once they enter the utility, can create correlations with the design that independent draws do not reproduce. When the correlation structure is known, a Gaussian copula keeps the marginal distributions of Step \ref{step:socio} and adds dependence. One draws $\zeta_n \sim \mathcal{N}(0, \Gamma)$, where $\Gamma$ is a correlation matrix, and sets $z_{nq} = F_q^{-1}(\Phi(\zeta_{nq}))$. Since $\Phi(\zeta_{nq}) \sim \mathcal{U}(0,1)$, each $z_{nq}$ keeps its marginal distribution $F_q$ by \eqref{eq:inverse}, while $\Gamma$ governs the dependence among the characteristics. This is a natural extension of the current version of the package.

With the design and the characteristics in place, utilities are computed and choices are made as in Steps \ref{step:utility} and \ref{step:choice}. The parameters are usually taken from a pilot study, an earlier study such as the case study, or the researcher's knowledge. The tool lets the user specify their mean vector and covariance matrix, so that random or deterministic, correlated or uncorrelated parameters can be introduced. For all respondents at once, \eqref{eq:beta} reads
\begin{equation}\label{eq:parmatrix}
B = \mathbf{1}_N \mu^{\top} + \Xi L^{\top},
\end{equation}
where row $n$ of the $N \times K$ matrix $B$ is $\beta_n^{\top}$, $\mathbf{1}_N$ is the vector of $N$ ones, and $\Xi$ is an $N \times K$ matrix of independent standard normal draws. A diagonal $\Sigma$ gives uncorrelated parameters and a full $\Sigma$ correlated ones, while $\Sigma = 0$ gives deterministic parameters, in which case the generating model is an MNL model by Proposition \ref{prop:dgp}.

\subsection{Performance Measures}\label{measures}

Because the parameters $\theta^0$ of the simulated data are known, the performance of an estimator can be measured exactly. Let the four steps be repeated $M$ times with independent random numbers, and let $\hat{\theta}^{(m)}_k$ and $\widehat{\mathrm{se}}^{(m)}_k$ be the estimate of component $k$ and its standard error in replication $m$. The bias, the RMSE, and the coverage of nominal 95\% confidence intervals are
\begin{align}
\mathrm{Bias}_k &= \frac{1}{M}\sum_{m=1}^{M} \left( \hat{\theta}^{(m)}_k - \theta^0_k \right), \mathrm{RMSE}_k = \left( \frac{1}{M}\sum_{m=1}^{M} \left( \hat{\theta}^{(m)}_k - \theta^0_k \right)^2 \right)^{1/2}, \label{eq:bias}\\
\mathrm{Coverage}_k &= \frac{1}{M}\sum_{m=1}^{M} \mathbf{1}\left\{ \left| \hat{\theta}^{(m)}_k - \theta^0_k \right| \leq 1.96 \widehat{\mathrm{se}}^{(m)}_k \right\}. \label{eq:coverage}
\end{align}
The RMSE decomposes as $\mathrm{RMSE}_k^2 = \mathrm{Bias}_k^2 + \frac{1}{M}\sum_{m}(\hat{\theta}^{(m)}_k - \bar{\theta}_k)^2$, where $\bar{\theta}_k$ is the average of the $M$ estimates, which separates systematic error from sampling variability. For a correctly specified model, Proposition \ref{prop:dgp} and the asymptotic normality of the estimator imply that, for large $N$, $R$, and $M$, the bias is close to zero, $\mathrm{RMSE}_k \approx ([I(\theta^0)^{-1}]_{kk}/N)^{1/2}$, and the coverage is close to 0.95. Comparing the simulated RMSE across sample sizes with this prediction gives an empirical counterpart of the sample size rule \eqref{eq:samplesize}. Departures from these benchmarks reveal finite-sample bias, simulation bias caused by a small $R$, or misspecification.

Since the case study targets the WTP, its recovery is also of interest. With a fixed price coefficient, the mean WTP for attribute $k$ is $-\mu_k/\beta_{\mathrm{Price}}$, as in \eqref{eq:wtpdist}, where $\mu_k$ is the mean of $\beta_{n,k}$ and equals $\beta_k$ if the coefficient is fixed. The estimate $\widehat{\mathrm{WTP}}_k = -\hat{\mu}_k/\hat{\beta}_{\mathrm{Price}}$ has, by the delta method, the approximate variance
\begin{equation}\label{eq:delta}
\operatorname{Var}\left(\widehat{\mathrm{WTP}}_k\right) \approx \frac{1}{\beta_{\mathrm{Price}}^2}\left( \Omega_{kk} - 2\frac{\mu_k}{\beta_{\mathrm{Price}}}\Omega_{kp} + \frac{\mu_k^2}{\beta_{\mathrm{Price}}^2}\Omega_{pp} \right),
\end{equation}
where $\Omega$ is the AVC matrix of the estimates, $k$ and $p$ index $\hat{\mu}_k$ and $\hat{\beta}_{\mathrm{Price}}$, and the right side is evaluated at the estimates in practice. The bias, RMSE, and coverage of the WTP follow from \eqref{eq:bias} and \eqref{eq:coverage} with $\theta^0_k$ replaced by the true WTP.

The simulator can also be confronted with real data collected with the same design. Let $p^{\mathrm{real}}_{sj}$ be the share of real respondents who choose alternative $j$ in situation $s$, and let $p^{\mathrm{sim}}_{sj}$ be the expected share under the simulator, which is positive because every logit kernel is positive and which can be estimated to any accuracy from a large simulated sample. The discrepancy between the two is measured by the average Kullback-Leibler divergence
\begin{equation}\label{eq:kl}
\Delta = \frac{1}{S}\sum_{s=1}^{S}\sum_{j \in \mathcal{C}} p^{\mathrm{real}}_{sj} \ln \frac{p^{\mathrm{real}}_{sj}}{p^{\mathrm{sim}}_{sj}},
\end{equation}
with the convention $0 \ln 0 = 0$. By Gibbs' inequality, $\Delta \geq 0$, with equality if and only if the real and simulated shares coincide in every situation. In a finite real sample $\Delta$ is almost always positive even when the simulator is correct. For a single situation answered by $N^{\mathrm{real}}$ respondents, $2N^{\mathrm{real}}$ times the inner sum of \eqref{eq:kl} is the likelihood ratio statistic of the simulated shares and is asymptotically $\chi^2$ with $J-1$ degrees of freedom under a correct simulator, but the panel structure makes the $S$ terms dependent, so their sum has no simple reference distribution. The observed value should therefore be compared with the distribution of $\Delta$ obtained when the real shares are replaced by the shares of independent simulated samples of the same size. This comparison acts as a parametric bootstrap test of the simulator.

\subsection{Sources of Discrepancy}\label{discrepancy}

Section \ref{sofscda} discussed several sources of bias, and the proposed simulator, still in an early stage of development, is not free of them. In the rose case, the scent of a rose may influence the choice. In many food purchases consumers taste the product before buying, and for other products they decide on the basis of information from brochures. Learning and experience are not simulated either, although earlier experiences influence later decisions in ways that vary among individuals and are difficult to simulate. Time also affects decision making, since compressed time intervals may lead to errors, but time pressure plays no role in the simulation. These effects can be stated in the notation of Section \ref{simulator}. A real respondent may behave according to
\begin{equation}\label{eq:real}
U^{\mathrm{real}}_{nsj} = \lambda_s \left( \beta_n + \delta_s \right)^{\top} x_{nsj} + \eta_{nsj} + \varepsilon_{nsj},
\end{equation}
where $\lambda_s > 0$ is the scale of utility in situation $s$, which falls under time pressure or fatigue and makes choices noisier, $\delta_s$ is a drift in tastes caused by learning along the sequence of situations, and $\eta_{nsj}$ collects sensory and informational factors, such as scent, taste, or brochure content, that are absent from the design. The simulator imposes $\lambda_s = 1$, $\delta_s = 0$, and $\eta_{nsj} = 0$, and it also sets the social influence parameter $\gamma$ of \eqref{eq:abm} to zero. When only the scale differs, with $\lambda_s = \lambda$ for all $s$, the scale argument following Lemma \ref{lem:logit} shows that real choices follow the same model with $\beta_n$ replaced by $\lambda\beta_n$. Estimates from real data then converge to $(\lambda\mu, \lambda^2\Sigma)$ instead of $(\mu, \Sigma)$, while the WTP, a ratio of coefficients, is unaffected. In the other cases the real and simulated choice probabilities differ, and \eqref{eq:kl} measures by how much.

\cite{burke1992comparing}, who compared dynamic consumer choice in real and computer-simulated environments, pointed out further discrepancies. Overall, creating a simulation environment that incorporates all these factors is a complex task that requires collaboration among researchers to extend the presented tool.

\section{Conclusion}\label{conclusion}

This paper has taken a multidisciplinary approach to bridge the gap between DCMs, experimental design, and simulation. It underscores the role of simulation in overcoming the practical constraints of real-world data collection, such as confidentiality concerns, time constraints, and high costs, and, more fundamentally, in providing data whose generating parameters are known.

\subsection{Key Contributions}

\subsubsection{Advancements in Discrete Choice Models}
DCMs, particularly RUMs, have a long history in the analysis of decision making across many fields. The paper restates their theoretical foundations in a common notation, from the logit form of the choice probabilities and the identification of the scale of utility to the MMNL model and its estimation by MSL, and it presents RRM models as an alternative behavioral framework. It also shows how the WTP, the central quantity in many applications, follows from these models as a ratio of coefficients that does not depend on the scale of utility.

\subsubsection{Simulation as a Tool for Data Generation}
The main contribution of the paper is a formal description of a simulation tool, implemented as an R package, that generates synthetic SC data for DCMs. Its data-generating process combines an experimental design, independently generated socioeconomic characteristics, normally distributed parameters, and Gumbel errors, and Proposition \ref{prop:dgp} shows that the resulting choices follow a panel MMNL model with known parameters. On this basis, the bias, RMSE, and coverage of the estimates and the recovery of the WTP measure the performance of an estimator, and the divergence \eqref{eq:kl} measures the distance between simulated and real choices. Researchers can therefore carry out detailed analyses, including the choice of design and sample size, before engaging in costly real-world experiments.

\subsubsection{Addressing Bias and Limitations}
The paper also examines the sources of bias and the limitations inherent in simulation-based studies. Simulations provide valuable approximations, but they cannot fully capture the complexity of human behavior. Emotional influences, social norms, bounded rationality, learning, and time pressure can separate simulated data from real-world data, and \eqref{eq:abm} and \eqref{eq:real} show how several of these factors can be expressed as social interactions, changes in scale, drifts in tastes, and omitted components of utility.

\subsection{Future Directions}
Several avenues remain for future research. First, integrating the RRM rule \eqref{eq:rrm} into the simulator, which only requires replacing the systematic utility by the negative regret \eqref{eq:regret}, would provide a more comprehensive tool for analyzing decision processes. Second, standardized methods for generating socioeconomic data, such as the Gaussian copula of Section \ref{casestudy}, would make the simulated populations more realistic. Third, the scale, drift, and social interaction terms of \eqref{eq:abm} and \eqref{eq:real} could be added to the simulator as options, so that their effect on the estimates can be studied in a controlled way. Finally, the tool can be used to compare the performance of DCMs with that of other classification models, such as artificial neural networks, offering a broader perspective on the analysis of consumer behavior.

\bibliography{references}

\begin{thebibliography}{10}
\expandafter\ifx\csname url\endcsname\relax
  \def\url#1{\texttt{#1}}\fi
\expandafter\ifx\csname urlprefix\endcsname\relax\def\urlprefix{URL }\fi
\expandafter\ifx\csname href\endcsname\relax
  \def\href#1#2{#2} \def\path#1{#1}\fi

\bibitem{ben1999discrete}
M.~Ben-Akiva, M.~Bierlaire, Discrete choice methods and their applications to short term travel decisions, in: Handbook of transportation science, Springer, 1999, pp. 5--33.

\bibitem{mcfadden1974utd}
D.~McFadden, \href{http://www.sciencedirect.com/science/article/pii/0047272774900036}{The measurement of urban travel demand}, Journal of Public Economics 3~(4) (1974) 303 -- 328.
\newblock \href {https://doi.org/https://doi.org/10.1016/0047-2727(74)90003-6} {\path{doi:https://doi.org/10.1016/0047-2727(74)90003-6}}.
\newline\urlprefix\url{http://www.sciencedirect.com/science/article/pii/0047272774900036}

\bibitem{mcfadden2001e}
D.~McFadden, \href{http://www.jstor.org/stable/2677869}{Economic choices}, The American Economic Review 91~(3) (2001) 351--378.
\newline\urlprefix\url{http://www.jstor.org/stable/2677869}

\bibitem{train2009dc}
K.~E. Train, Discrete choice methods with simulation, Cambridge university press, 2009.

\bibitem{chorus2008random}
C.~G. Chorus, T.~A. Arentze, H.~J. Timmermans, A random regret-minimization model of travel choice, Transportation Research Part B: Methodological 42~(1) (2008) 1--18.

\bibitem{chorus2010new}
C.~G. Chorus, A new model of random regret minimization, European Journal of Transport and Infrastructure Research 10~(2) (2010).

\bibitem{trainmcfadden}
D.~McFadden, K.~Train, Mixed mnl models for discrete response, Journal of Applied Econometrics 15~(5) (2000) 447--470.

\bibitem{bunch1996comparison}
D.~S. Bunch, J.~J. Louviere, D.~Anderson, A comparison of experimental design strategies for multinomial logit models: The case of generic attributes (1996).

\bibitem{hess2014handbook}
S.~Hess, A.~Daly, Handbook of choice modelling, Edward Elgar Publishing, 2014.

\bibitem{kirk2012experimental}
R.~E. Kirk, Experimental design, Handbook of Psychology, Second Edition 2 (2012).

\bibitem{kuhfeld2003marketing}
W.~F. Kuhfeld, Marketing Research Methods in SAS., Citeseer, 2003.

\bibitem{rose2008designing}
J.~M. Rose, M.~C. Bliemer, D.~A. Hensher, A.~T. Collins, Designing efficient stated choice experiments in the presence of reference alternatives, Transportation Research Part B: Methodological 42~(4) (2008) 395--406.

\bibitem{carlsson2010design}
F.~Carlsson, Design of stated preference surveys: Is there more to learn from behavioral economics?, Environmental and Resource Economics 46~(2) (2010) 167--177.

\bibitem{kahneman2003maps}
D.~Kahneman, Maps of bounded rationality: Psychology for behavioral economics, American economic review 93~(5) (2003) 1449--1475.

\bibitem{michaud2013willingness}
C.~Michaud, D.~Llerena, I.~Joly, Willingness to pay for environmental attributes of non-food agricultural products: a real choice experiment, European Review of Agricultural Economics 40~(2) (2013) 313--329.

\bibitem{hensher2013random}
D.~A. Hensher, W.~H. Greene, C.~G. Chorus, Random regret minimization or random utility maximization: an exploratory analysis in the context of automobile fuel choice, Journal of Advanced Transportation 47~(7) (2013) 667--678.

\bibitem{gusarov2020exploration}
N.~Gusarov, A.~Talebijamalabad, I.~Joly, Exploration of model performances in the presence of heterogeneous preferences and random effects utilities awareness, Da2Pl conference (2020).

\bibitem{chorus2014random}
C.~Chorus, S.~van Cranenburgh, T.~Dekker, Random regret minimization for consumer choice modeling: Assessment of empirical evidence, Journal of Business Research 67~(11) (2014) 2428--2436.

\bibitem{masiero2019understanding}
L.~Masiero, Y.~Yang, R.~T. Qiu, Understanding hotel location preference of customers: comparing random utility and random regret decision rules, Tourism Management 73 (2019) 83--93.

\bibitem{sharma2019park}
B.~Sharma, M.~Hickman, N.~Nassir, Park-and-ride lot choice model using random utility maximization and random regret minimization, Transportation 46 (2019) 217--232.

\bibitem{mao2020does}
B.~Mao, C.~Ao, J.~Wang, B.~Sun, L.~Xu, Does regret matter in public choices for air quality improvement policies? a comparison of regret-based and utility-based discrete choice modelling, Journal of cleaner production 254 (2020) 120052.

\bibitem{iraganaboina2021evaluating}
N.~C. Iraganaboina, T.~Bhowmik, S.~Yasmin, N.~Eluru, M.~A. Abdel-Aty, Evaluating the influence of information provision (when and how) on route choice preferences of road users in greater orlando: Application of a regret minimization approach, Transportation Research Part C: Emerging Technologies 122 (2021) 102923.

\bibitem{wong2020revealed}
S.~D. Wong, C.~G. Chorus, S.~A. Shaheen, J.~L. Walker, A revealed preference methodology to evaluate regret minimization with challenging choice sets: a wildfire evacuation case study, Travel Behaviour and Society 20 (2020) 331--347.

\bibitem{hillel2021systematic}
T.~Hillel, M.~Bierlaire, M.~Z. Elshafie, Y.~Jin, A systematic review of machine learning classification methodologies for modelling passenger mode choice, Journal of choice modelling 38 (2021) 100221.

\bibitem{soleymani_forecasting_2024}
S.~Soleymani, A.~Talebi, Forecasting {Solar} {Irradiance} with {Geographical} {Considerations}: {Integrating} {Feature} {Selection} and {Learning} {Algorithms}, Asian Journal of Social Science and Management Technology 6~(1) (2024) 85--93.

\bibitem{talebi2024integrating}
A.~Talebi, S.~P. Haeri~Boroujeni, A.~Razi, Integrating random regret minimization-based discrete choice models with mixed integer linear programming for revenue optimization, Iran Journal of Computer Science (2024) 1--15.

\bibitem{rose2008stated}
J.~M. Rose, M.~C. Bliemer, Stated preference experimental design strategies, Handbook of transport modelling (2008) 151--180.

\bibitem{street2007construction}
D.~J. Street, L.~Burgess, The construction of optimal stated choice experiments: Theory and methods, Vol. 647, John Wiley \& Sons, 2007.

\bibitem{bliemer2011experimental}
M.~C. Bliemer, J.~M. Rose, Experimental design influences on stated choice outputs: an empirical study in air travel choice, Transportation Research Part A: Policy and Practice 45~(1) (2011) 63--79.

\bibitem{rose2009constructing}
J.~M. Rose, M.~C. Bliemer, Constructing efficient stated choice experimental designs, Transport Reviews 29~(5) (2009) 587--617.

\bibitem{bliemer2009efficient}
M.~C. Bliemer, J.~M. Rose, D.~A. Hensher, Efficient stated choice experiments for estimating nested logit models, Transportation Research Part B: Methodological 43~(1) (2009) 19--35.

\bibitem{bliemer2010construction}
M.~C. Bliemer, J.~M. Rose, Construction of experimental designs for mixed logit models allowing for correlation across choice observations, Transportation Research Part B: Methodological 44~(6) (2010) 720--734.

\bibitem{street2005quick}
D.~J. Street, L.~Burgess, J.~J. Louviere, Quick and easy choice sets: constructing optimal and nearly optimal stated choice experiments, International journal of research in marketing 22~(4) (2005) 459--470.

\bibitem{voytekhovsky2001fedorov}
Y.~L. Voytekhovsky, The fedorov algorithm revised, Acta Crystallographica Section A: Foundations of Crystallography 57~(4) (2001) 475--477.

\bibitem{bonabeau2002agent}
E.~Bonabeau, Agent-based modeling: Methods and techniques for simulating human systems, Proceedings of the national academy of sciences 99~(suppl 3) (2002) 7280--7287.

\bibitem{gilbert2019agent}
N.~Gilbert, Agent-based models, Vol. 153, Sage Publications, Incorporated, 2019.

\bibitem{brock2003mcsi}
W.~Brock, S.~Durlauf, \href{https://EconPapers.repec.org/RePEc:nbr:nberte:0288}{Multinomial choice with social interactions}, NBER Technical Working Papers 0288, National Bureau of Economic Research, Inc (2003).
\newline\urlprefix\url{https://EconPapers.repec.org/RePEc:nbr:nberte:0288}

\bibitem{zhang2019agent}
N.~Zhang, X.~Zheng, Agent-based simulation of consumer purchase behaviour based on quality, price and promotion, Enterprise Information Systems 13~(10) (2019) 1427--1441.

\bibitem{zhang2007agent}
T.~Zhang, D.~Zhang, Agent-based simulation of consumer purchase decision-making and the decoy effect, Journal of business research 60~(8) (2007) 912--922.

\bibitem{lockshin2006using}
L.~Lockshin, W.~Jarvis, F.~d’Hauteville, J.-P. Perrouty, Using simulations from discrete choice experiments to measure consumer sensitivity to brand, region, price, and awards in wine choice, Food quality and preference 17~(3-4) (2006) 166--178.

\bibitem{munizaga2005testing}
M.~A. Munizaga, R.~Alvarez-Daziano, Testing mixed logit and probit models by simulation, Transportation Research Record 1921~(1) (2005) 53--62.

\bibitem{wendel2020designing}
S.~Wendel, Designing for behavior change: Applying psychology and behavioral economics, " O'Reilly Media, Inc.", 2020.

\bibitem{Amirreza2019}
A.~Talebijamalabad, N.~Gusarov, I.~Joly, \href{https://github.com/Amirreza-96/sdcm}{{sdcm}: {S}imulation of {D}iscrete {C}hoice {M}odels}, Grenoble INP, Grenoble, France, version 0.0.0.9000 (2020).
\newline\urlprefix\url{https://github.com/Amirreza-96/sdcm}

\bibitem{white1982maximum}
H.~White, Maximum likelihood estimation of misspecified models, Econometrica 50~(1) (1982) 1--25.

\bibitem{gromping2018r}
U.~Gr{\"o}mping, R package doe. base for factorial experiments, Journal of Statistical Software 85~(1) (2018) 1--41.

\bibitem{gronmping2014r}
U.~Gr{\"o}nmping, R package frf2 for creating and analyzing fractional factorial 2-level designs, Journal of Statistical Software 56~(1) (2014) 1--56.

\bibitem{traetsgenerating}
F.~Traets, D.~G. Sanchez, M.~Vandebroek, Generating optimal designs for discrete choice experiments in r: The idefix package (2019).

\bibitem{burke1992comparing}
R.~R. Burke, B.~A. Harlam, B.~E. Kahn, L.~M. Lodish, Comparing dynamic consumer choice in real and computer-simulated environments, Journal of Consumer research 19~(1) (1992) 71--82.

\end{thebibliography}

\end{document}